\documentclass[sigconf]{acmart}
\usepackage{listings}
\usepackage{amsmath}
\begin{document}

\title{Omelets Need Onions \\ \large E-graphs Modulo Theories via Bottom-up E-Matching}


\author{Philip Zucker}
\email{philzuckerblog@gmail.com}

\begin{abstract}
    E-graphs\cite{eggpaper} are a data structure for equational reasoning and optimization over ground terms. One of
     the benefits of e-graph rewriting is that it can declaratively handle useful but difficult to orient identities like 
     associativity and commutativity (AC) in a generic way. However, using these generic mechanisms
      is more computationally expensive than using bespoke routines on terms containing sets, multi-sets, linear expressions, polynomials, and binders.
        A natural question arises: How can one combine the generic capabilities of e-graph rewriting with these specialized theories?
        This paper discusses a pragmatic approach to this e-graphs modulo theories (EMT) question using two key ideas: bottom-up e-matching and semantic e-ids.


\end{abstract}


\maketitle

\section{Introduction}

\subsection{Terms Modulo Theories}

It is instructive to dial back from the complexity of e-graphs to see how one can make sense of the "baking in" of theories for ordinary tree-like terms.

The most basic form of a term is a function symbol and an ordered list of children
\begin{lstlisting}[language=Caml]
type term = Fn of string * term list
\end{lstlisting}

We can extend this to a more general form of term with specialized containers of children

\begin{lstlisting}[language=Caml]
type term = 
  | Fn of string * term list
  | MS of string * term multiset
  | Set of string * term set
  | Poly of string * term poly
  | LitInt of int
\end{lstlisting}





The specialized data structures like set and multiset can be implemented using regular lists. Structurally canonical list representatives can be found by using sorting and deduplication \cite{hashingmodtheories}. A more performant thing to do can be to use Patricia tries, which are also structurally canonical but support faster set and map operations.

Separately, one can also use smart constructors \cite{smartconstructor} to apply baked in rewrite rules immediately upon construction. They check all rules which have the smart constructor symbol at the head of the pattern. If the rewrite rules are convergent, one can maintain the invariant the terms are only created in canonical form. 

If the terms are stored in a structurally canonical way, one can use regular hash consing \cite{hashconsing} to intern them. This supports fast equality checks modulo the theories via a pointer check.

This shows that equality checking and interning terms modulo sufficiently nice theories is not that hard, but what about pattern matching?



 
 








\subsection{What patterns are possible?} \label{patternpossible}

Pattern matching and unification are a logical and syntactic slant on the notion of equation solving \cite{traatbook} \cite{BAADER2001445}. The problems are formulated as goal equations $\{p_1 =^? t_1, p_2 =^? t_2, ...\}$ containing variables $X, Y,...$ to be solved, and the solutions are substitutions $\sigma$ , which are mappings from variables to terms. Pattern matching and unification operate as backwards reasoning methods that breaks down the goal equation set until it is refined into a substitution.

"E-matching" in the e-graphs community has a connotation of matching a pattern against an e-graph, but in the term rewriting community it has a connotation of solving a pattern matching problem modulo an equational theory $E$ consisting of a set of universally quantified equality axioms. The e-graph version of e-matching is related to the term version of e-matching with $E$ being the theory of ground equations that the e-graph represents.


Equational theories fall into different classes of how many substitutions they may require for a complete solution of their E-unification or E-matching problems. Some theories will only have a finite set of solutions, some a countably infinite set, and others an uncountably infinite set.

\begin{itemize}

\item First order syntactic pattern matching is equation solving for a simple structural theory of terms with ordered children. It is unusual in the spectrum of e-matchers in that it either succeeds once or fails. For example, the goal $cons(X,Y) =^? cons(zero,nil) $ is solved by the substitution $\{X \mapsto zero, Y \mapsto nil\}$ and the goal $cons(X,X) =^? cons(zero,nil)$ has no solution.


\item Another case may be that a finite number of solutions must be returned. This is the case for some structural theories like AC symbols, sets and multisets. The goal $\{X , Y\} =^? \{1, 2\}$ is solved by $\{X \mapsto 1, Y \mapsto 2\}$ and $\{X \mapsto 2, Y \mapsto 1\}$. 



\item An enumerably infinite number of solutions is possible. $X + Y =^? 42$ considered over the integers has solutions $\{ \{X \mapsto n, Y \mapsto 42 - n  \} | n \in \mathbb{Z} \}$. $X \cap Y =^? \{1,2,3\}$ has solutions $\{ \{X \mapsto A, Y \mapsto B\} | \{1,2,3\} \subseteq A,B \} $ .



\item Finally, even this may not be possible for some theories. Consider solving $X + Y =^? \pi$ over $\mathbb{R}$. The only remaining possibility is to instead return a constraint, which more or less just returns the frozen pattern.

\end{itemize}
In short, if solutions need to be returned as a collection of substitutions, matching on terms with baked in theories runs the spectrum from expensive to hopeless.









\section {Bottom Up E-Matching Plays Nicer With Theories}

One way to win is to refuse the play the game.

In a typical statement of the pattern matching problem, one is given a  pattern $p$ to be matched against a particular term $t$. 

A different but related problem is to instead be given a term bank $T$ which is scanned in an attempt to fill each pattern variable $V$ in $p$ \cite{bottomup}. When the problem is posed in this way, the set of substitutions found is bounded as a subset of $\{\{ v \mapsto t | v \in V \} | t \in T \}$. This is the bottom-up e-matching problem. 

It is important to note that while an e-graph can be described as a compressed representation of a set of ground equations $E$, really it is more like the pair of this set of equations and a termbank $(E, T)$. For example, we may consider the set of equations $\{foo(a) = b, c = c\}$ to be the same as the equation set $\{foo(a) = b\}$ since the equation $c = c$ has no interesting equational content. Nevertheless, if we assert these two sets into the e-graph, we get different e-graphs, one which has the an e-class and e-node for $c$ in its termbank $T$ and one which does not. The ability to seed the e-graph termbank with useful terms is crucial for equality saturation.

Naive equality saturation has a built in implicit methodology for expanding the termbank. At every stage, it adds to the termbank all subterms that are equal to something already in the termbank by a rewrite rule. This is a useful heuristic for adding terms that may be useful via congruence closure, but it is incomplete for equational search  \cite{remybirkhoff}. Other rules to expand the termbank need to be considered \cite{GeMoura2009}.


This python code shows an example of a shallow embedding of bottom-up e-matching. It scans choices of variables in the form of \lstinline{for} loops. This is so simple it does not really need a library and can instead be expressed as an idiom.

\begin{lstlisting}[language=Python]
  # A shallow bottom-up embedding of the rewrite
  # foo(bar(X), Y) -> biz(X)
  for X in eclasses:
      for Y in eclasses:
          try:
              # possibly failing lookup in hashcons
              lhs = foo[bar[X], Y]  
              # construct rhs
              rhs = biz(X)          
              # set equal
              egraph.union(lhs,rhs) 
          except e:
              pass
  \end{lstlisting}

This is not an apples and oranges comparison to the complexity of a virtual machine implementation of top-down e-matching \cite{z3ematching} since top-down e-matching specialized to a particular pattern can be written as loops as well, one loop for every e-class expansion. Subjectively speaking, the resulting loops are harder to read, because the entirety of the pattern never appears in one syntactic location.



For a rough asymptotic complexity comparison of the algorithms, consider an e-graph with $E$ e-classes and $N$ e-nodes. On average, each e-class has $N/E$ e-nodes.

Top-down e-matching pays a factor $E$ to even get started for a scan of a root e-class to pattern match over. Every time top-down e-matching expands an e-class into its e-nodes, it pays a multiplicative factor $\frac{N}{E}$ due to this search. Because of this, top-down e-matching is exponential in cost in terms of the depth of the pattern $d$. The cost of top-down e-matching is something like $O(E(\frac{N}{E})^d)$.

Bottom-up e-matching pays for a scan of the e-classes $E$ for each of the number of variables $V$ in the pattern. Constructing the instantiated pattern is just a lookup and pays the cost of a hash table or indexing tree lookup, which is typically something like $O(\ln(N))$. The cost is therefore something more akin to $O(E^V d \ln(N) )$.

For different patterns or termbanks, top-down or bottom-up may be more asymptotically performant.

A bad pattern for top-down e-matching would be a deep one with few variables. For example, 

$$foo(foo(foo(foo(foo(foo(X))))))$$.

A bad pattern for bottom-up e-matching would be a shallow one with many variables. For example,

$$foo(bar(X), bar(Y), bar(Z), bar(W), bar(V))$$


A reason that top-down matching may feel natural is that is a method that makes sense for pattern matching against a pointer chasing implementation of single term. This is the typical situation in many programming language-like contexts, but not the only choice of representation.

Another reason bottom-up e-matching may not feel natural is that the most commonly considered termbank is one consisting of ground syntactic terms with no theories. In this case $\frac{N}{E} = 1$ and the exponential character of top-down matching is gone. It is a dominant method. I believe that people do not necessarily use simple syntactic terms because they do not want baked in theories. They do it because baking in theories is complex and expensive. Bottom-up e-matching ameliorates these issues.

Bottom-up e-matching can be seen as a particular query plan for relational e-matching using a generic join \cite{wang2023freejoinunifyingworstcase}. Likewise top-down e-matching can be seen as the same. 

The positive perspective of relational e-matching \cite{zhang2022relationalematching} is that it gives one freedom to find a good efficient query plan. If you have the implementation budget, the best plan will sometimes include top-down, bottom-up, and side-to-side moves. 

A negative perspective of relational e-matching is that it doesn't commit to a choice, leaving behind a vague gesturing to the new problem of query optimization which is yet another subsystem to design and implement. Adding theories to the other directions besides bottom-up requires extra interfaces and implementation work, whereas adding theories to bottom-up comes nearly for free, only requiring the already existent lookups and canonizers. Bottom-up e-matching sits on an interesting point in the Pareto frontier of simplicity, features, and performance.




Bottom-up e-matching has the killer feature is that it proceeds in the same direction as adding terms into the e-graph. Therefore it uses the same indexing structures and canonization that already exist. Bottom-up e-matching grounds pattern variables as quickly as possible, and ground terms are the e-graph's bread and butter. Ground is good.

Top-down e-matching requires the extra ability to expand from e-classes to e-nodes and from e-nodes to their children. This is, as we have seen in section \ref{patternpossible}, extra challenging or even impossible in the face of theories. Top-down keeps the pattern in an ungrounded state as long as possible. Unground is bad.

Bottom-up e-matching can emulate the results of top-down e-matching. If it is possible to enumerate the subterms of your target term, you can fill a term bank with these subterms. Bottom-up e-matching will then find the same matches top-down will. In the presence of some theories enumerating subterms is not possible or meaningful and top-down e-matching will not work at all. You can still bottom-up match over a term databank. In this sense, bottom-up e-matching is more powerful than top-down e-matching.



The maximally naive bottom-up e-matching that scans over all e-classes can be improved by filtering on the intersection of all the tables where the variable occurs and by pushing intermediate constructions upwards in the loop ordering. 

\begin{lstlisting}[language=python]
for X in set(bar.keys()).intersection(biz.keys()):
  try:
    t1 = bar[X]
    for Y in set([k[1] for k in foo.keys()]):
      try:
          lhs = foo[t1, Y]
          rhs = biz(X)
          egraph.union(lhs,rhs)
\end{lstlisting}

One can also prune the loops by modeling using multi-sorted logic. One only needs to scan over the e-ids corresponding to the sort of the pattern variable. This can be achieved by having separate union finds per sort. Which brings us into our next topic.

\section {Generalized Union Finds and Semantic E-ids}

After bottom-up e-matching, the second key ingredient for e-graphs modulo theories is to generalize the e-id and the union find \cite{grobneregraph}. E-ids are generalized to structured e-ids / semantic e-ids / values. The union find becomes a theory specific canonizer that supports equality assertions.

A first way to see this might make sense is to note that a simple e-graph can be seen as a uninterpreted partial-function minimal model $\mathcal{M}$ of the equations $t_1 = t_2$ and term existence assertions $\downarrow t$ that produced it. The collection of e-ids $\mathcal{V}$ corresponds to the set of values in the model and the e-node tables correspond to the data of partial functions $\lBrack f \rBrack$ of the model. In the simple e-graph, the uninterpreted domain of values has no structure except equality. One can instead consider interpretations onto other semantic or structured domains like polynomials, linear expressions, terms, constructors, sets, multisets and so on. These corresponds to structured e-ids. Multiple kinds of structured e-ids can be most simply supported by using multi-sorted logic. A static type check assures that expressions of different sorts are never equated. The role of mediating between these sorts is performed by the e-nodes and congruence closure through e-nodes.

Another point to note is that even in hash consing for terms modulo theories, there is utility in having both interned and non-interned forms of terms \cite{kontroliefficient}. A rewriting smart constructor should not bother interning intermediate expressions until the fully canonized form has been reached. The e-nodes are akin to the interned forms and the structured e-ids are more akin to the non-interned forms. The structured e-ids are in a sense kept in extracted non-interned term-like form for easy deep inspection and theory specific destructive manipulation. While the full interning of hash consing makes for the simplest and fastest equality checks, this is a knob that can be turned between complete structural traversal and a simple pointer comparison. 

Another way to see that this generalized union find idea makes sense is to ask "what interface does the union find present?". The internals of the union find and the integer representation of the e-ids are irrelevant to how they are used in the e-graph. Really one only needs at minimum the abstract interface

\begin{lstlisting}[language=Caml]
type t
type eid
val create : unit -> t
val eq     : t -> eid -> eid -> bool
val fresh  : t -> eid (* makeset *)
val canon  : t -> eid -> eid (* find *)
val assert_eq : t -> eid -> eid -> unit (* union *)
\end{lstlisting}

This API is largely the same as that required by the theory of an SMT solver minus the support for backtracking. This complies with the schematic equation $EMT = SMT - SAT$. It is particularly close to the interface of Shostak style theory combination \cite{shostaklight}. 

\subsection{Union Finds solve Ground Atomic Equations}

The union find can be seen as a canonizer of a system of atomic equations. The Knuth Bendix completion of a ground atomic equational theory is guaranteed to terminate and the resulting rewrite system can be interpreted as a union find \cite{groundegraph2024}. 

Consider for example the atomic ground equation system 
 \label{groundeq}
\begin{equation}
\begin{split}
e_1 = e_2 \\
e_2 = e_3  \\
e_2 = e_1  \\
e_3 = e_1  \\
e_4 = e_5  \\
\end{split}
\end{equation}

Running Knuth Bendix completion with a term ordering $$e_i < e_j \triangleq i < j$$ results in an atomic ground rewrite system

\begin{equation}
\begin{split}
e_2 \rightarrow e_1 \\
e_3 \rightarrow e_1 \\
e_5 \rightarrow e_4
\end{split}
\end{equation}

The $\rightarrow$ can be represented by a parent pointer in a union find.



\subsection{Example Theories}

However, atomic ground theories are not the only ones that are guaranteed to be completable to canonizers. There is a spectrum of theory specific "union finds" that support canonization and ground equality assertion \cite{MARCHE1996253}. Many of them have the flavor of a specialized theory specific ground Knuth Bendix completion.

\begin{itemize}

\item  Linear equations can be put into row echelon form by Gaussian elimination. The row echelon form can be used to quotient a vector space by a subspace. A set of linear equations can be put into the matrix form $Ax=0$. For example, each equation of the system in \ref{groundeq} can be encoded as a row of a matrix. 

\[
\begin{bmatrix}
1 & - 1 & 0 & 0 & 0 \\
0 & 1 & -1 & 0 & 0 \\
-1 & 1 & 0 & 0 & 0 \\
-1 & 0 & 1 & 0 & 0 \\
0 & 0 & 0 & 1 & -1 \\
\end{bmatrix}
\begin{bmatrix}
   e_1 \\
   e_2 \\
   e_3 \\
   e_4 \\
   e_5 \\
\end{bmatrix}
 = 0
\]

The row echelon form corresponds to a ground rewrite system. 

\[
\begin{bmatrix}
1 & - 1 & 0 & 0 & 0 \\
0 & 1 & -1 & 0 & 0 \\
0 & 0 & 0 & 1 & -1 \\
\end{bmatrix}
\begin{bmatrix}
   e_1 \\
   e_2 \\
   e_3 \\
   e_4 \\
   e_5 \\
\end{bmatrix}
 = 0
\]

It can be used starting from the bottom of the matrix to canonically normalize $\sum_i a_i e_i $ expressions where $a_i \in \mathbb{Q}$. For example $e_1 + 3e_3 + 5e_5 \rightarrow e_1 + e_3 + 5e_4 \rightarrow e_1 + 3e_2 + 5 e_4 \rightarrow 4e_1 + 5e_4$. The method however works just as well with non-atomic asserted equations like $4e_1 + 5e_4 = 13e_2 + 2e_3$, extending the capabilities of the union find to baked in linear expressions.



From the e-graph as model perspective, the semantic domain of these semantic e-ids is a vector space of uninterpreted e-ids quotiented by the subspace spanned by the asserted linear equations. $\mathcal{V} = \mathrm{span}_\mathbb{Q} \{ e_1, e_2, e_3, ... \} \Big/ V_E$. Affine equations work similarly.

\item Gr\"{o}bner bases are the generalization of row echelon form to multivariate polynomial equation systems \cite{Cox2015}\cite{grobnerpap}. They are guaranteed to exist, but much more expensive to compute than a row echelon form. An example application is trigonometry and Euclidean geometry which can be put into a form that is essentially polynomial, for example trigonometric identities like $\sin(x)^2 + \cos(x)^2 = 1$.   The semantic values are the quotient ring of polynomials modulo the ideal generated by the polynomial equations. $\mathcal{V} = \mathbb{Q}[e_1,e_2,..] \Big/ I_E$
\item Ground multiset equations are also completeable. The structured e-ids are finitely supported multisets $\mathbb{N}^{\{e_1, e_2, ...\}}$. This gives each sort one special associative and commutative operation. \cite{acegraph} This can be seen as a special case of Gr\"{o}bner basis calculation as each monomial $x^n y^m z^k...$ can be seen as a multiset of it's variables with the multiplicity given by the exponent. Indeed, the termination of Buchberger's algorithm is guaranteed by multiset orderings on the produced polynomials. It can also be seen as integer lattice rewriting via Hilbert or Graver bases and has connections to the theory of integer programming \cite{sturmfels1996grobner}. Another way of formulating the semantic values is as a quotient of the semi-module generated by the simple e-ids $\mathcal{V} = \mathrm{span}_\mathbb{N} \{ e_1, e_2, e_3, ... \} \Big/ V_E$. 

\item In an amusing self reflection, the semantic e-id could themselves be ground terms. The theory solver could be an e-graph, which is a canonizer for ground term equations. One could separate functions symbols to those considered outer and inner. This would enable for example, two e-graph implementations to interact with each other. Similarly, it is possible to use terms of a external SMT solver as structured e-ids \cite{extz3egraph}. Canonization can be achieve by black box SMT queries.

\item A union find can be built that carries group elements on it's edges \cite{Frühwirth_2009} \cite{lemerre2024labeled} \cite{groupoidunion} for a theory of uninterpreted e-ids modulo a group action.  The composition law of the group enables composition while chasing up the parent pointers of the union find, and the group inverse enables turning comparison of leaf-to-root to root-to-leaf. In this manner it can be used to store an equivalence relation modulo a group action. For example the group of integer addition $x + 6 = y - 3$ while subsumed by the row echelon method can be more simply and efficiently implemented using this technique. Permutation groups are also very useful in the context of nominal rewriting and slotted e-graphs \cite{slotted}.

\item Atomic completion modulo constructors and primitives can be seen as a method to merge the notion of e-ids and primitives. The semantic domain is the disjoint union of uninterpreted symbols and the set of primitive $ \mathcal{V} = \{ e_i \}  \uplus \mathcal{P} / E$. E-class ids may either be uninterpreted or grounded to a concrete entity like an integer, string, or constructor. Tie breaking of the union operation on the union find has to be adjusted such that if a primitive is unioned, it is always the root. Unioning two compatible constructors can be resolved by syntactic unification. Unioning two incompatible constructors is considered an error because you are asserting an impossibility. Constructors and injectivity are already supported in a simple e-graph and rules, but this is a form of baking them in. Having the constructor term materialized makes occurs checks and bisimilarity compaction for coalgebraic datatypes possible to perform by ordinary term handling routines rather than some traversal of the e-graph. This is related to the ideas of co-egraphs \cite{coegraph} which add a co-function/observation table alongside e-node function table. The identification of e-ids with values suggests a method to support lambda terms by using closure values as structured e-ids in an intriguing connection with the ethos of Normalization by Evaluation. In NbE, the semantic domain is extended by syntactic forms representing stuck terms.


\item A Boolean theory of ground clauses $ \lnot a \lor b \lor c$ or ground Horn clauses $b \land c \implies a$ can be canonized by ordered resolution. This can also be extended to ground superposition clauses $ \lnot(x = y) \lor z = x \lor c = true$. This may give a method to encoding contextual equality. One can also similarly consider using a theory of boolean rings.  

\item One not entirely satisfying way of baking in convergent destructive rewrite rules is if they operate over a disjoint signature from the e-nodes of the egraph and encoding terms using those constants from those signatures into the structured e-ids rather than in e-nodes. Fuller support (at cost) is to use Knuth Bendix completion as suggested below.

\end{itemize}

It is possible to use theory solvers that may not terminate during their rebuilding / completion steps. One may either hope they do, or terminate their completion process early and accept the incomplete canonizers. The e-graph typically has a self-healing character in the face of incomplete canonization.

\begin{itemize}
\item String completion gives a baked in theory of associativity. The semantic domain is the finitely presented monoids $<e_1, e_2, ... | R_E>$. The completion of string rewriting in general will not succeed because string rewriting is rich enough to encode Turing complete computation \cite{traatbook}. Baked in sequences are particularly compelling for modelling stateful computation. Similar theories include finitely presented groups, non-commutative rings, and differential algebra.

\item One may choose e-ids to be so structured as to be ground terms themselves. Unfailing Knuth Bendix completion is how destructive rewriting (demodulation) can be used while retaining completeness in equational theorem proving. A given set of desirable destructive rewrite rules may be completed ahead of time, but adding in new ground equations in general generates new critical pairs, requiring the completion process to resume during rebuilding of the "union find".

\end{itemize}


\section{Alternative Approaches to EMT}

The above approach is not the only one that might reaosnably go under the moniker e-graphs modulo theories.

Koehler et al describe an "extraction-based substitution" \cite{koehler2022sketchguidedequalitysaturationscaling} method for dealing with substitution of lambda terms. One extracts a term, simplifies it using bespoke routines, and reinserts it into the e-graph. This method works for all rewrite theories for which one wants to avoid intermediate junk, not just lambda terms. Some of this junk may be useful though, such as in AC where one wants to bubble certain term together rather than just have a single flattened sorted form.

Another approach is to mirror all e-graph unions and assertions into a side-car SMT solver \cite{extz3egraph}. This enables SMT based guards on rules, for example the theory of linear inequalities. This technique has analogs in the form of Constraint Logic Programming \cite{JAFFAR1994503}, constrained Horn clauses, or constrained rewriting  \cite{constrainedrewrite}.

A third approach that is "off by one" from semantic-eids is to merge the concept of containers and e-nodes. Similiarly as was shown for terms, e-nodes may have a set, multiset, or polynomial of children. Bottom-up e-matching still works, and top-down can somewhat be made to work by extending the signature of to return multiple results.

\begin{lstlisting}
fn children(self) -> Vec<Vec<Id>>
\end{lstlisting}

\section{Related Work}

Other examples of e-graph extensions that are evocative of the structured e-id idea are colored e-graphs \cite{singher2023colored} which tag e-ids with a notion of context and slotted e-graphs \cite{slotted} which have some of the flavor of both generalized e-ids and generalized e-nodes. 

There have been a number of extensions to rewriting to term rewriting modulo theories. Chapter 11 of Baader and Nipkow has a survey of some options \cite{traatbook}. Of particular interest is Normalizing rewriting \cite{MARCHE1996253} which has the most similarity to the approach here.

String Knuth Bendix systems appear in computational group theory applications. These systems by construction have built in associativity axioms.

The Vampire \cite{vampire} and E theorem provers \cite{eprover} are two powerful automated theorem provers with some built in support for AC symbols.

There is a line of work for congruence modulo AC applied in the Alt-Ergo SMT solver \cite{conchon12lmcs}

The Maude system has support for efficient AC rewriting \cite{maudeac}.


Relational AC matching \cite{relationalac} uses the ideas of Relational E-Matching \cite{zhang2022relationalematching} to encode AC matching into database queries.

\section{Conclusion}

Bottoms up! Go forth and saturate!


\begin{acks}
I'd especially like to thank Cody Roux for many many belligerant grillings about term rewriting. I'd also like to thank Max Willsey, Yihong Zhang, Yisu Remy Wang, Graham Leach-Krouse, Greg Sullivan, Caleb Heibling, Sam Lasser, and Ryan Wisnesky for helpful discussions.
\end{acks}

\bibliographystyle{ACM-Reference-Format}
\bibliography{refs}

\end{document}